# Interplay between magnetic properties and thermoelectricity in misfit and Na cobaltates

J. Bobroff,[1] S. Hébert,[2] G. Lang,[1] P. Mendels,[1] D. Pelloquin,[2] A. Maignan[2]

[1] Laboratoire de Physique des Solides, Univ. Paris-Sud, UMR8502, CNRS, F-91405 Orsay Cedex, France
[2] Laboratoire CRISMAT, UMR 6508, CNRS et EnsiCaen, 14050 Caen (France)

We present a comparative study of $CoO_2$ layers in the Bi-misfit and $Na_xCoO_2$ cobaltates. Co NMR measures the *intrinsic* susceptibility of the Co layers and is not affected by spurious contributions. At low dopings where room-temperature thermopower (TEP) is large, Curie-Weiss susceptibilities are observed in both materials. But NMR and µSR experiments find neither charge nor spin order down to low temperatures in Bi-misfit, in contrast to the case of $Na_xCoO_2$. This demonstrates that metallicity, charge and magnetic orders are specific of the Na layers in $Na_xCoO_2$ whereas strong correlations are generic of the cobaltates physics and could explain the large TEP.

High efficiency of thermoelectric devices is requested for opening the way to large scale applications such as thermoelectric refrigerators or power sources. This relies upon a large figure of merit $ZT = S^2T/\rho\kappa$, where S, the Seebeck coefficient, measures the thermoelectric power (TEP), ρ is the resistivity, κ is the thermal conductivity and T the temperature. Combining a large *S* together with a low resistivity is usually impossible to achieve in standard metals or semiconductors [1]. Many alternatives have been investigated to circumvent this limitation, from rattling skutterudites semiconductors to narrow band heavy fermions. The recent discovery of unexpectedly large *ZT* in the Cobaltates $Na_xCoO_2$ has opened a new route [2]. There, orbital or spin degrees of freedom together with strong correlations might be used as an independent source of entropy resulting in large TEP together with good conductivity [2,3,4]. However, the origin of such large TEP is still highly debated and not necessarily linked to correlations. It could be caused by specific spin or charge orders known to occur in the Co layers [5,6] or accounted for by a more conventional metallic picture [7]. Understanding the origin of the correlations in this promising compound and their interplay with transport and thermoelectric properties is a key element to settle this issue. This would help to decide whether low dimensional strongly correlated oxides could play a role in the future developments of thermoelectricity.

In this frame, we focus on the Bi-misfit cobaltates, which display large TEP as well [8,9]. Bi-Misfits and $Na_xCoO_2$ feature identical $CoO_2$ layers but differ by their dopant layer. In $Na_xCoO_2$, it consists of a single Na layer where the Na can order structurally. In Bi-misfit, it is composed of a thick rock-salt (RS) layer incommensurate with the Co layers [8], and the doping is varied by changing its cationic and oxygen composition. This results in a smoother spatial variation of the Coulomb potential due to the dopant layer than in $Na_xCoO_2$, because of the RS structure, incommensurability and thickness. By comparing the properties of Bi and Na cobaltates, we are able to distinguish the intrinsic properties of the Co layers from those due to the dopant layer nature. The intrinsic spin susceptibility measured by NMR shows a flat metallic-like behavior at high doping (low x) and a Curie-Weiss behavior at low doping ($x \geq 0.65$), revealing the presence of correlations like in $Na_xCoO_2$. However, using NMR and µSR, we do not find any charge or magnetic orders down to low temperatures. Furthermore, Bi-misfit cobaltates are found much less conducting than Na cobaltates. This comparison demonstrates that large TEP in these oxides are due to strong correlations.

We prepared powder samples of the 4-layer Bi-family $[Bi_2M_2O_4]^{RS}.[CoO_2]_m$ (M=Ba,Sr,Ca and m the misfit ratio) named hereafter BiMCoO, using the procedure detailed in [8]. By changing both the cation M and the oxygen content through thermal treatments, we are able to span the phase diagram in an unprecedented large range of doping, equivalent to the domain $x \approx 0.6 \to 0.9$ in $Na_xCoO_2$. To access higher TEP and lower dopings than previous studies, BiCaCoO powders were also synthesized in a sealed ampoule with primary vacuum, leading to S(*T*=300K)=210 µV/K, to be compared to the air value S=150 µV/K. Annealing this sample at *T*=400°C in oxygen pressure ($PO_2$ = 100 Bars) leads to an intermediate value S=190 µV/K. This variation is likely due to a change of the oxygen content, hence a variation of the $CoO_2$ layer doping [10]. All structures were checked using X-ray and electron diffraction coupled to EDS analyses. A four probe technique and a steady-state method were respectively used to measure resistivity and TEP in a Physical

Properties Measurements System, using ceramic bars with ultrasonically deposited indium contacts. The μSR measurements were performed at P.S.I (Switzerland) facility. The NMR measurements were carried out in a variable field magnet at fixed frequency $\nu_{rf}$=56.3MHz. Fourier-Transform echo recombination was used to record each spectrum, with a delay between pulses of 8 μs and a repetition time of 10 msec. The NMR integrated intensities of the various samples were identical within 20%. This ensures that all the Co nuclei are detected in the reported spectra. Isotropic shifts were evaluated using the centre of gravity of the spectra, taking for the $^{59}$Co gyromagnetic ratio $\gamma/2\pi$=10.053MHz/Tesla.

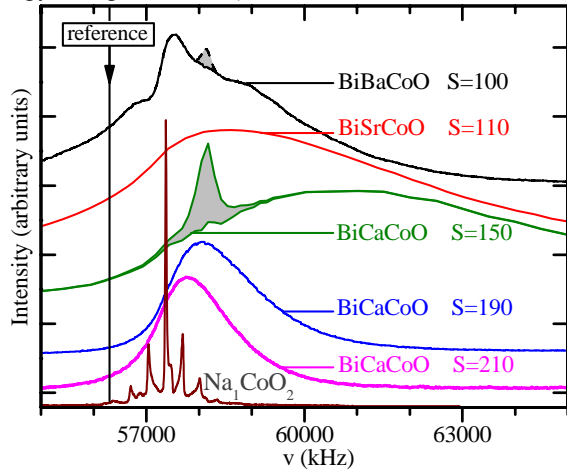

fig.1 : Plot of Co NMR spectra measured at $T$=5K for various misfit compositions together with the undoped $Na_1CoO_2$ [11]. Related TEP measured at T=300K in μV/K are indicated. In some cases, an additional $Co_3O_4$ spurious contribution is observed (grey shaded area).

Typical Co NMR spectra are shown in fig.1 at $T$=5K. A broad line is observed, together in some cases with a narrow peak due to the presence of about 10% of spurious $Co_3O_4$ [12]. This explains why one cannot rely on macroscopic susceptibility measurements in these misfits: the $Co_3O_4$ paramagnetic contribution is too large to be safely subtracted. On the contrary, with a local probe such as NMR, it is straightforward to measure the intrinsic $CoO_2$ susceptibility using the shift of the main line. As compared to the spectrum of non oriented powder of $Na_1CoO_2$, the misfit spectra are broader, more shifted, and do not show any set of satellite lines. In $Na_1CoO_2$, these quadrupolar satellites originate from the effect of the electric field gradient (EFG) at Co site [11]. In misfits, the incommensurability of the nearby RS layers strongly distributes the Co EFG, which could explain the absence of any quadrupolar structure. In addition, as we measure non-oriented powders, the large anisotropy of the hyperfine fields which occurs in doped $CoO_2$ layers [13] leads to a distribution of the shift, i.e. an additional broadening. The sharp difference with $Na_1CoO_2$ where all Co are 3+ low spin S=0 state implies that no isolated $Co^{3+}$ ion is present in misfits on the timescale of NMR (~ 10 μs). We performed contrast measurements similar to those done in $Na_xCoO_2$ [13] to probe the existence of different Co sites, by changing the delay between pulses in the NMR echo sequence. The whole line displays a homogeneous relaxation, implying that *a unique Co valence state* is detected, in contrast with the charge segregation observed in $Na_{0.67}CoO_2$.

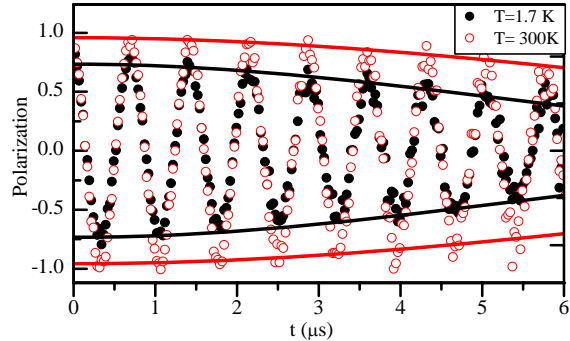

fig.2 : μSR polarization in a 100 Gauss transverse field for BiCaCoO (S=210 μV/K) shows an oscillating behavior typical of a paramagnetic regime. The initial decrease with decreasing temperature originates from spurious contributions while BiCaCoO remains non magnetic.

If any magnetic order were present, the spectra would split, broaden or wipe-out. This is not the case, which indicates the absence of spin order or freezing down to $T$=5K in all compounds. We further checked this possibility by performing μSR measurements in the three BiCaCoO samples. As shown in fig. 2, at $T$=300K, a weak applied transverse field makes most of the polarization of the muons oscillate. This implies that no large internal field is present in the compound where muons were implanted. The weak Gaussian relaxation of the polarization (solid lines) reflects the existence of a small distribution of fields of a few *G* mainly due to the nuclei dipolar fields. When decreasing temperature, about 10 to 20% of the muons spin do not oscillate anymore, because these muons belong to spurious phases such as CoO or $Co_3O_4$ which order at $T_N$=289K and 34K. In addition, the Gaussian envelope of the oscillating part narrows a little, implying the apparition of a small additional source of relaxation. If this were due to a spin ordering or freezing, we evaluated that it would correspond to moments of maximum 0.012 $\mu_B$ per Cobalt in the worse and unlikely case where muons would all stop as far as possible from Co layers. Measurements in zero field lead to similar conclusions. The two other BiCaCoO samples show similar behavior. In another μSR study, Sugiyama *et al.* found no order as well down to $T$=1.8K and 2.3K for BiBaCoO and BiSrCoO respectively [6]. But they argued that BiCaCoO showed spin freezing, in contrast to our findings. However, only 12% of their sample volume actually ordered, which is not significant. They further claimed that a universal dome-shaped relation between $T_N$ and S occurred in all cobaltates. Our study performed on a *single* family and on a large range of S values rules out the universality of such a dome phase diagram for $T_N$. This unambiguously demonstrates that *magnetic ordering is not responsible or linked to the existence of a large TEP in cobaltates*.

This absence of order sharply contrasts with the case of Na-cobaltates which become antiferromagnets at $T_N$ = 19K, 22K and 28K [14], as shown in fig.4.c. Incommensurability of misfits should play no role since BiBaCoO is commensurate. Therefore, either the much smaller interlayer distance in $Na_xCoO_2$ (~5 Å) than in misfits (~15 Å) or the Na crystallographic order may explain the antiferromagnetism. In the latter case, such orders could be induced by a Co spatial charge ordering or by a reconstruction of the Fermi Surface.

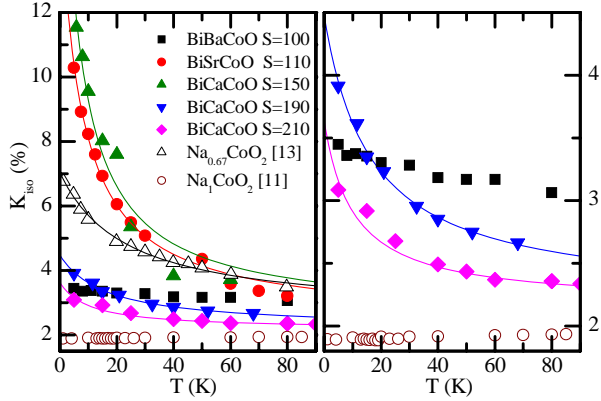

fig.3 : $^{59}$Co NMR isotropic shift $K_{iso}$ of the misfit compounds shown on two different scales (left and right panel). At low S, $K_{iso}$ is flat, while it shows a Curie-Weiss behavior at higher S (full lines are Curie-Weiss fits).

We now focus on the paramagnetic regime. The isotropic part of the shift of the NMR line, $K_{iso}$, is proportional to the magnetic susceptibility of Co. This shift is the only reliable way to determine the intrinsic susceptibility here as already stressed above. As expected, $K_{iso}$ is always higher and more *T*-dependent than that of the band insulator $Na_1CoO_2$, as shown in fig.3. In the metallic BiBaCoO with low TEP, a flat behavior is observed, reminiscent of a metallic Pauli susceptibility. We measured a similar behavior in another low TEP metallic TlSrCoO misfit whose structure and macroscopic properties are reported in [15]. In contrast, in BiSrCoO and BiCaCoO which are less metallic and have a higher S, the shift, hence the susceptibility, follows a Curie-Weiss behavior (solid lines in fig.2):

$$K_{iso} = \frac{A_{hf} C}{T + \Theta} + K_{orb}^{iso}$$

where $A_{hf}$ is the hyperfine coupling and $K_{orb}^{iso}$ is the *T*-independent orbital shift reported in fig.4.a. The Curie Constant C first increases sharply with increasing S, then decreases again. The Weiss temperature $\Theta$ is always lower than 100K, but hard to determine exactly from data taken on powders.

For a quantitative comparison with $Na_xCoO_2$, we first need to evaluate the misfit Co layer doping. In $Na_xCoO_2$, the Seebeck coefficient S at *T*=300K scales almost linearly with x for $0.5 \leq x \leq 0.98$ as seen in fig.4.e [16]. We use this arbitrary scaling to deduce an effective x doping in our misfit samples where we measured the TEP as well [17]. From our measurement of S at *T*=300K, we find x=0.66(5), 0.69(5), 0.71(5), 0.79(3), 0.86(3) and 0.89(3) for TlSrCoO, BiBaCoO, BiSrCoO, and the three different BiCaCoO. This increase of x when decreasing the ion size (Ba→Sr→Ca) hence decreasing the misfit parameter m (2→1.82→1.67) is consistent with electroneutrality. It is also the same as the one determined using the Fermi Surface area for BiBaCoO [18]. It is then possible to plot the various properties of misfit and Na-cobaltates on a single x scale in fig.4 where we also included results we obtained on TlSrCoO. All families fall on a *unique* line for $K_{orb}^{iso}$ (fig.4.a). This strongly supports our phenomenological evaluation of x for misfits. The linear dependence of $K_{orb}^{iso}$ with Co valence could be linked to the crystal field splitting of the $t_{2g}$-$e_g$ levels and its filling, as advocated in [13]. The T-dependence of the shift is evaluated in fig.4.b. using $\Delta K = K_{iso}(T=5K)-K_{iso}(T=80K)$, a quantity which evolves as the Curie constant C, and which can be estimated for flat behavior as well. Misfit and Na cobaltates both evolve from a flat susceptibility to a Curie-Weiss dependence with increasing x. The very similar susceptibilities observed for both Bi and Na families at $x \approx 0.7$ in fig.3 and 4 implies that neither the interlayer distance nor the ordering of Na layers has a strong influence on the magnetic susceptibility at this doping. For $x \geq 0.75$, no reliable $Na_xCoO_2$ susceptibility data is available, and the present study allows us to probe this region for the first time. The susceptibility still displays a Curie-Weiss behavior, but C decreases with x. The Curie-Weiss temperature $\Theta$ also decreases from 50-100K to 0–20K within our limited accuracy. In a localized picture, higher x means a reduction of the number of spins and C, and an increase of the distance between spins, i.e. a decrease of $\Theta$, as demonstrated in a t-J model [4]. It is not obvious that $Na_xCoO_2$ would show similar dependences, since the Na Coulomb potential has been argued to induce strong correlations even at low doping [5].

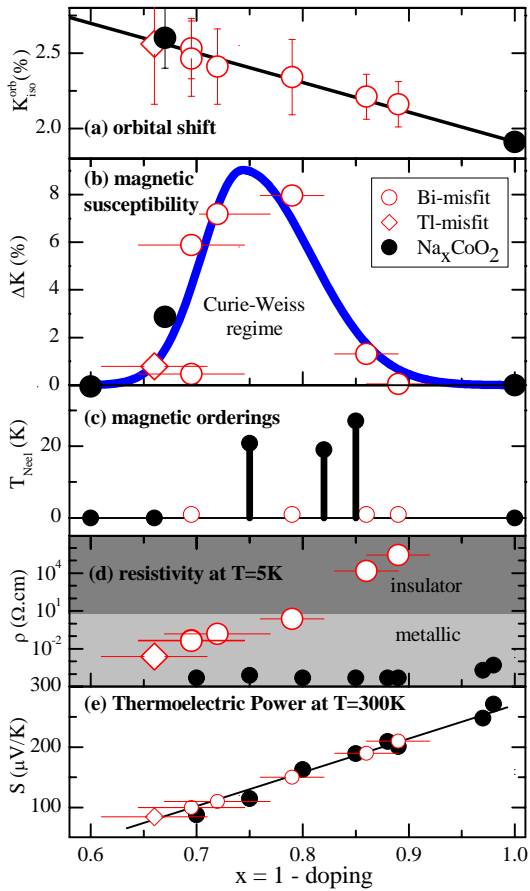

fig.4 : Phase diagram of misfit and Na cobaltates. (a) Isotropic orbital shift measured at $T$=300K (b) The $T$-dependence of the misfit magnetic susceptibility is evaluated through $\Delta K=K_{iso}(5K)-K_{iso}(80K)$ and compared to published results for $Na_xCoO_2$ [11,13]. Blue line is a guide to the eye. (c) $T_N$ in $Na_xCoO_2$ [14] compared to the upper bound found for any order for misfit from this study and [6]. (d),(e) ρ at $T$=5K and S at $T$=300K are compared to $Na_xCoO_2$ single crystal data of [16].

We finally report on fig.4.d and 4.e. resistivity and TEP. The misfits evolve from an insulator to a metal behavior when increasing doping, i.e. decreasing x, as confirmed also by the $T$-dependences of the resistivity [10]. It is not surprising to find a metal-insulator transition when varying hole doping (1-x). But one cannot understand along the same line why Na cobaltates are good metals at all dopings. Recent ARPES measurements reveal as well qualitative differences in the dispersion curves for $x \geq 0.7$ between cobaltates and misfits [18]. Perhaps specific arrangements of the Na allow for better hoppings or the RS incommensurability in misfit is detrimental to transport through random potential scattering of the conduction electrons.

We address now the question of the TEP origin. While magnetic orders, charge orders and good conductivity appear specific to $Na_xCoO_2$, Curie-Weiss susceptibilities and large TEP are found both in Bi-misfit and Na cobaltates and remain down to low dopings. We conclude that TEP is likely linked to the strong correlations within the Co layers, as proposed in [3,4,19] rather than the band properties in a more conventional metallic picture [7]. Understanding the nature of this link and how correlations result in the properties summarized in fig.4 is an open and challenging issue. It is not only appealing on the fundamental side but also essential for thermoelectric applications using these strongly correlated materials in the future.

We acknowledge H. Alloul, I. Mukhamedshin, V. Brouet and P. Limelette for fruitful discussions and A. Amato for assistance at μSR facility. This work was supported by the ANR "OxyFonda" and by the EC FP 6 program, Contract No. RII3-CT-2003-505925.